\documentclass[12pt]{iopart}

\usepackage{iopams}
\usepackage{caption}
\usepackage{graphicx}
\usepackage[T1]{fontenc}       % Use modern font encodings

\usepackage[binary-units=true]{siunitx}
\usepackage{array,multirow}
\usepackage[square,sort,comma,numbers]{natbib}
\usepackage[table]{xcolor}
\usepackage{hyperref}

\begin{document}
	
	\newcolumntype{P}[1]{>{\centering\arraybackslash}p{#1}}
	\definecolor{lightgray}{RGB}{200,200,200}
	\newcommand{\vect}[1]{\boldsymbol{\bi{#1}}}
	\newcommand{\rvec}{\vect{r}}
	\newcommand*{\doi}[1]{\href{http://dx.doi.org/#1}{doi: #1}}

\title[Predicting bioaccumulation using molecular theory]{Predicting bioaccumulation using molecular theory: A machine learning approach}

\author{Sergey Sosnin$^1$,Maksim Misin$^2$,Maxim V. Fedorov$^1$}
\address{$^1$Skolkovo Institute of Science and Technology, Skolkovo Innovation Center, Moscow 143026, Russia}
\address{$^2$Institute of Chemistry, University of Tartu, Ravila 14a, Tartu 50411, Estonia}

\ead{m.fedorov@skoltech.ru}

	\begin{abstract}
		In this work, we present a new method for predicting bioaccumulation factor of organic molecules. The approach utilizes 3D convolutional neural network (ActivNet4) that uses solvent spatial distributions around solutes as input. These spatial distibutions are obtained by a molecular theory called three-dimensional reference interaction site model (3D-RISM).We have shown that the method allows one to achieve a good accuracy of prediction.
		Our research demonstrates that combination of molecular theories with modern machine learning approaches can be effectively used for predicting properties that are otherwise inaccessible to purely physics-based models.
	\end{abstract}

\maketitle

	\section{Introduction}

	Molecular theories such as three-dimensional reference interaction site model (3D-RISM) \cite{Beglov1997vrd,Hirata2003tpg,Ratkova2015teb} or molecular density functional theory (MDFT) \cite{Jeanmairet2013tow,Ramirez2005vwr,Gendre2009wap} rely on statistical mechanics derived approximations to estimate the equilibrium distribution of solvent around solvated molecules. In turn, these distributions can be related to the solution properties of solvated molecules \cite{Hansen2013wia,Ben-naim2006vpu}. Examples include solvation free energy\cite{Du2000uyo,Palmer2010upn,Misin2015wdu}, partial molar volume \cite{Ratkova2015teb,Misin2017uha}, salting out constants \cite{Misin2016vee}, binding free energies \cite{Genheden2010vvf,Gussregen2017uuu,Sugita2016vcs} and others. However, using purely theoretical approach, it is practically impossible to relate these distributions to the substance's biological effects, such as toxicity or bioaccumulation.

	The above does not mean that solvation structure is not useful for the understanding influence of chemical compounds on the living organisms. On the contrary, the rich information encoded in the solvation shell can be used to understand whether a given compound is hydrophobic or hydrophilic \cite{Lum1999val}, guess if it will be able to pass certain membrane channels \cite{Roux1991tlc} or in case solvent contains ions, estimate its affinity towards them \cite{Misin2016vee}. All this information is directly related to compound biological effects but can not be expressed rigorously using equations. On the other hand, the machine learning methods are usually quite good at finding and quantifying these relations.

	In this article, we utilize a 3D convolutional neural network (CNN) to estimate bioaccumulation factor in a number of organic molecules. As an input, we use three-dimensional distributions of water around these molecules, obtained using 3D-RISM with Kovalenko-Hirata closure (KH)\cite{Kovalenko1999usb}. Artificial neural networks (ANNs) have been previously used for predicting biological effects of organic molecules \cite{FPBasedQSAR2012,KinaseQsar2013,Merck2015}. However, they were combined with a very broad set of descriptors that have diverse physical meanings. Here we focus on a single property, solvation shell structure, and attempt to establish its utility. To determine whether a 3D-RISM procedure is necessary we also train electrostatic potential based model and compare the results.
	
	\section{Theoretical Background}
	
	\subsection{Representation of Molecules}
	
	In chemical informatics molecules are typically represented using a variety of descriptors. The simplest descriptors are derived from a graph representation of a molecule. Although this is a very compact representation which requires little storage, it ignores a lot of spatial details and can not be used as an input for the majority of neural networks.
	
	The numerical descriptors based on: fragmental\cite{FragmentalReview2011}, topological\cite{TopologicalReview2013}, quantum\cite{QuantumReview1996}, physicochemical\cite{PhysicochemicalReview2004} and other properties are more suitable for machine learning approaches and have been used extensively in traditional QSAR/QSPR models. However, typically, accurate models require combining of descriptions of different nature, that makes the interpretation of such models difficult.
	
	An interesting approach to these limitations is offered by 3D QSAR approaches that can combine a rich numerical description of a molecule with relatively easily interpretable models. One of the most widely used 3D QSAR method is \textit{Comparative Molecular Field Analysis (CoMFA)}\cite{CoMFA1988}. In this method an interaction energy between the molecule and a probe atom \cite{3DQSAR1993} is measured for each 3D grid point forming 3D molecular fields. These 3D fields are analysed by a partial least-squares (PLS)\cite{PLS2001} regression to build a predictive model. This method is implemented in the popular commercial software \textit{Sybyl-X}\cite{SYBYLX2017} and has been widely used in science and pharmaceutical companies in the last 25 years.
	
	The majority of molecules have a number of conformers with different 3D representations. There is no guarantee that low energy conformers obtained from \textit{in-silico} experiments would correspond to the active conformation of a ligand \cite{ComparativeQSARStudies2017}. To overcome this limitation a number of alignment-free methods have been proposed\cite{RecentAdvancesInQSAR2010}, and a batch of multi-dimensional QSAR methodologies have been developed\cite{4DQSARINDD2010}, but there is no universal approach. This problem has inspired us to search for a method that can handle multiple conformations of a molecule. 
	
	\subsection{3D Reference Interaction Site Model (3D-RISM)}
	
	Calculation of an equilibrium distribution of solvent around an arbitrary molecule is a challenging problem in molecular modeling \cite{Ratkova2015teb}. It can be done using molecular dynamics simulation, but extremely long simulation times are needed to obtain smooth solvent distributions \cite{Luchko2010uhh}. MDFT method\cite{Gendre2009wap}, proposed by Borgis and ER-Theory \cite{Matubayasi2000uqh} developed by Matubayasi and Nakahara can be applied to this problem. As an alternative to the molecular simulations, molecular solvation theories (that are also refereed to as integral equation theories) offer a less computationally expensive way of calculating the 3D solvation structure around a molecule \cite{Chandler1986tqr,Ratkova2015teb}.
	One of these methods, a three-dimensional reference interaction site model (3D-RISM) became very popular for calculations of the distribution of solvent sites (atoms) around the molecular solute\cite{Ratkova2015teb,Kovalenko1999usb,Kovalenko2000vzf}. 
	
	As a result of 3D-RISM calculations, one obtains a single-site \textit{density distribution function} (local density) $\rho_{\gamma}(\vect{r})$ of the solvent site $\gamma$ around the solute molecule. We used a 3-point model of water (SPC/E) meaning that the calculation produces density distributions for both oxygen and hydrogen atoms. These density distributions can be regarded as a variant of molecular fields that we discussed in the previous section. Notice that the densities obtained from RISM calculations are not exact \cite{Hirata2003tpg,Misin2017uha}, but can be successfully used to predict variety of both chemical and biological properties using either semi-empirical corrections \cite{Truchon2014vty,Misin2015wdu,Misin2016tqn,Misin2016vee,Misin2016tqn} or QSPR approaches \cite{Palmer2015vaf}.

	The 3D-RISM main equation can be written as \cite{Hirata2003tpg,Misin2017uha}:
	$$
	h_{\gamma}(\rvec) = \sum_{\alpha=1}^{n_s} (\chi_{\alpha\gamma} * c_{\alpha}) (\rvec),
	$$
	where $*$ denotes convolution, $n_s$ stands for the number of solvent sites, and $h{\gamma}(\rvec) = \rho_{\gamma}(\rvec)/\rho_{\gamma} - 1$, usually refered to as the total correlation function, is a function introduced for convenience. $c(\rvec)$ is a direct correlation function that can be defined via $c_\gamma(\rvec)=-\mu_{\gamma}^{\ast}(\rvec)/(kT)$. Here $k$ is the Boltzmann constant, $T$ is temperature, and $\mu_{\gamma}^{\ast}(\rvec)$ is an excess chemical potential (relative to ideal gas) of the solvent site $\gamma$. Finally, $\chi_{\alpha\gamma}(r)$ is a site-site susceptibility function that can be obtained from a bulk solvent radial distribution functions. More conveniently, $ \chi_{\alpha\gamma}$ can be calculated from a separate 1D-RISM calculation\cite{DRISM1992,Ratkova2015teb}.
	
	The above equation is usally coupled with a separate closure relation that provides another connection between $h_{\gamma}(\rvec)$ and $c_{\alpha}(\rvec)$. In this work we used a simple, but robust \textit{Kovalenko-Hirata (KH)}\cite{KHClosureOrig2000} closure:
	$$
	h_{\gamma}(\rvec) + 1 =
	\cases{
	\exp\left[-\beta u_{\gamma}(\rvec) + h_{\gamma}(\rvec) - c_{\gamma}(\rvec)\right],&\text{if $h(\rvec)\leq0$;}\\
	1 -\beta u_{\gamma}(\rvec) + h_{\gamma}(\rvec) - c_{\gamma}(\rvec),&\text{if $h(\rvec)>0$;}
	}
	$$
	where $\beta=1/(kT)$ and $u_{\gamma}(\rvec)$ is a potential energy between the solvent site $\gamma$ and the solute molecule. Together the above systems of equations are usually iteratively solved until both $h_{\gamma}(\rvec)$ and $c_{\alpha}) (\rvec)$ achieve the predefined convergence criteria. 
	
	\subsection{Bioconcentration factor (BCF)}
	
	In this work, we built a model for predicting the bioconcentration factor, BCF (more specifically, we predicted its decimal logarithm $\mathrm{log_{10}BCF}$). This factor is the ratio between the concentration of an organic compound in biota and in water:\cite{BCFCanada2003} 
	\begin{equation}
	BCF = \frac{Concentration_{biota}}{Concentration_{water}} \label{eqn:bcf}
	\end{equation}
	This factor is an important parameter for estimating the potential danger of an organic compound. It is one of the parameters that determine the labeling of the compound under \textit{Registration, Evaluation, Authorisation and Restriction of Chemicals (REACH)} program in European Union. The ability of a compound to penetrate and conserve in an organism may influence the toxicity and mutagenicity of the compound, and so may reveal potential environmental risks. Generally, if a compound has BCF value of more than 5000 (or $\mathrm{\log_{10} BCF} > 3.67$), it is regarded as potentially dangerous. There are several methods to measure and estimate the confidence of the BCF data, described in details in ref. \citenum{Arnot2006}. The detailed description of BCF estimation is published in Organisation for Economic Cooperation and Development (OECD) guideline No 305\cite{OECD3052012}. It should be textitasized that determining of BCF in in-vivo experiments is a very expensive procedure. 
	
	Over the years, several models for BCF prediction have been published. Arnot and Gobas have proposed a linear model that predicted BCF as a function of the uptake and elimination of an organic compound by an aquatic organism. Since BCF is related to logP and water solubility\cite{Arnot2006}, some authors proposed models that utilise these descriptors \cite{LinearBCFModel2007}. These linear models work satisfactory only for moderately hydrophobic compounds, but fail to address strongly hydrophobic chemicals\cite{BCFMolDescriptors2005}. Additionally, \textit{LogP} is a parameter that must be measured separately and this may be problematic. Another notable model has been produced by Zhao et al.\cite{BCFHybrid2008} using a hybrid of a number of machine learning methods. Their model managed to produce an impressive accuracy ($R^2=0.8\mathrm{,  RMSE}=0.59$), albeit on a somewhat curated dataset.
	
	To conclude the above, modeling of the bioconcentration factor is an important research area due to the difficulties associated with its experimental evaluation and importance of such models for regulatory purposes. 
	
	\section{Methods and Materials}
	\textit{Databases}
	We used the dataset collected by USA Environmental Protection Agency for their T.E.S.T. QSAR platform for risk estimation\cite{TESTEPA2016}.
	This dataset has been split into training and test subsets in the same manner as it was done by US EPA, and statistical values on the test set are published. We used them as a baseline for our model. There are 541 molecules in the training set and 135 molecules in the test set.   
	
	We used rkdit\cite{RdKit2017}, an open-source cheminformatic toolkit, to perform basic molecular routines and to calculate the geometries of molecules.
	
	\textit{Conformers Generations}
	For deep neural networks, high amount of diverse data is a key factor to success. Our approach to conformer generation and selection is similar to the method from the article\cite{RdKitConformersGeneration2015} and is briefly described below.
	
	At the first stage of the algorithm, we generate a number of conformers by rotating the bonds of a molecule. This is followed by an energy minimization step, consisting of 5000 iterations and performed using the universal force field (UFF) \cite{UFF1992}. Subsequently, all conformers with RMSD (computed on the heavy atoms) of less than \SI{0.5}{\angstrom} are discarded. If the number of conformers exceeds the pre-defined limit, then the post-processing procedure from paper\cite{RdKitConformersGeneration2015} is performed (we discuss this procedure in more details in the Supporting Information).

	\textit{Force field assignment}
	We used AmberTools16\cite{AmberTools162016} package to calculate the partial charges of each molecule using AM1-BCC\cite{AM1BCC2002} semi-empirical model. At this stage, for some molecules from the training set the calculations have not converged, and these molecules were eliminated. These partial charges were used for further 3D-RISM calculations and for the evaluation of electric potential fields. 
	
	\textit{Electric potential field}
	We calculated the electrostatic potential around every molecule by placing it into a cube box (voxel grid) with the dimensions of $\mathrm{\SI{35}{\angstrom}\times \SI{35}{\angstrom} \times \SI{35}{\angstrom}}$ using a grid with the step size of \SI{0.5}{\angstrom} for all dimensions. For each grid point $j$ in the $70\times70\times70$ box we calculated 
	$$ 
	U_{j} = \frac{1}{4\pi\epsilon_0} \sum_{i}^{atoms}\frac{q_{i}}{|\rvec_{ij}|} 
	$$ 
	Where $\epsilon_0$ stands for the vacuum permittivity, $ q_{i} $ is a partial charge of the $i$-th solute atom and $|\rvec_{ij}|$ is the distance between the atom $i$ and the grid point $j$.
	
	We implemented the code for the potential calculation on GPU units using framework \textit{cupy} (a part of \textit{chainer} framework\cite{Chainer2015}), and used it for ``on-the-fly'' 3D fields calculation. 
	Note that before being fed to the neural network, all values of the potential were scaled by a constant factor to scale it into a reasonable range. 

	\textit{Lennard-Jones potential field}
	We calculated the Lennard-Jones potential field in the similar manner to electric potential described above. For each atom pairs $ \epsilon $ and $ \sigma $  parameters have been taken and mixed with a probe atom in accordance with \textit{GAFF2} forcefield\cite{AmberTools162016} mixing rules. We used oxygen bonded with hydrogen as probing atom (noted as \textit{``oh''} in \textit{AmberTools}) to make the modeling routine close to potential estimation with water as a solvent in RISM. 
	
	$$ 
	U_{j} = 4\epsilon_{ij}*\sum_{i}^{atoms}({(\frac{\sigma_{ij}}{|\rvec_{ij}|})}^{12} - {(\frac{\sigma_{ij}}{|\rvec_{ij}|})}^{6})
	$$ 
	We clipped values of energies more than 2 \textit{kcal/mol} to precise 2 \textit{kcal/mol}. and less than zero to zero to avoid numerical dispersion. It was established in our research that smoothing Lennard-Jones potentials is a crucial step, and we transformed the values by $ e^{-U} $ in every grid point.
	
	\textit{3D-RISM Calculations}
	3D-RISM equation was solved using \textit{rism3d.snglpnt} program from AmberTools16\cite{AmberTools162016} package. Water with temperature 298 K was used as a solvent. Site-site susceptibility functions of bulk water ($\chi_{\alpha\gamma}(\rvec)$) were calculated using DRISM method by \textit{drism} program from the same package. For 3D-RISM we used a $\mathrm{\SI{35}{\angstrom}\times \SI{35}{\angstrom} \times \SI{35}{\angstrom}}$ grid with \SI{0.5}{\angstrom} step size. The resulting oxygen and hydrogen density distributions were saved as HDF5\cite{HDF52017} binary files. We ran a separate 3D-RISM calculation for each conformer. If more than 50\% of 3D-RISM calculations did not converge, the molecule was marked as ``failed'' and was eliminated from the dataset. 
	
	\begin{figure}
		\begin{centering}
		\includegraphics[width=.30\linewidth]{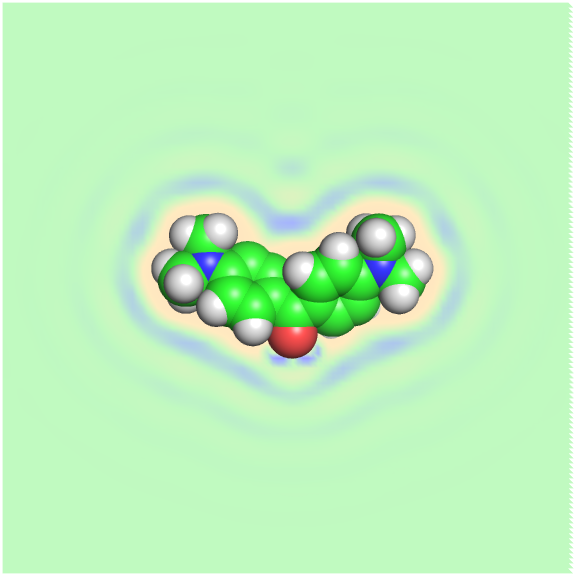}
		\includegraphics[width=.30\linewidth]{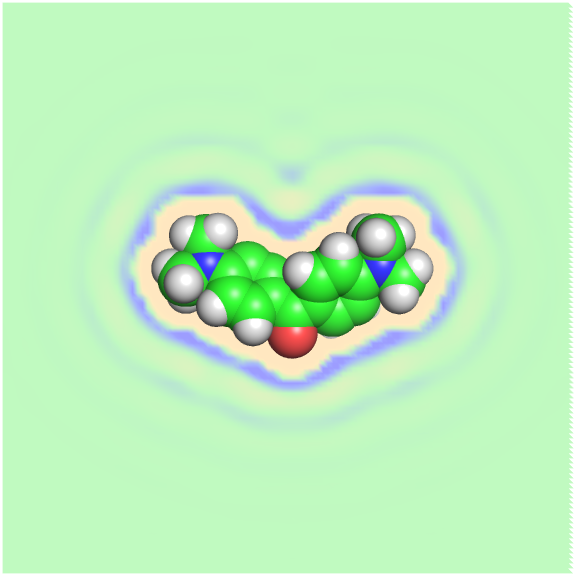}
		\includegraphics[width=.30\linewidth]{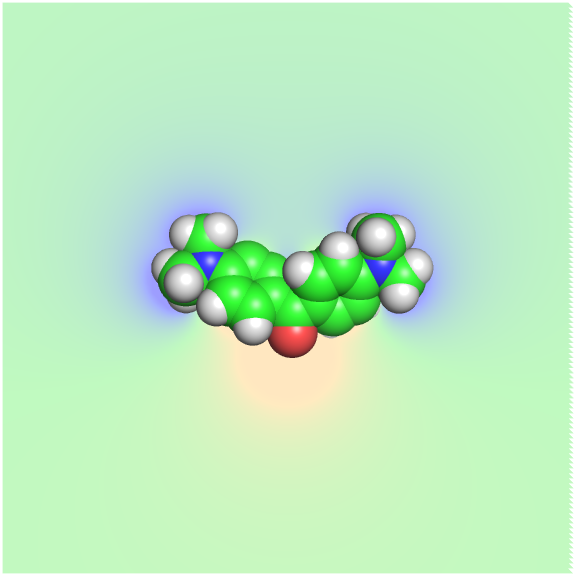}
		\end{centering}
		\caption{An example of the visualization of the scalar fields for a molecule as 2D slices taken by the principal axis (Left -- a visualization of hydrogens density. Center -- a visualization of oxygen density. Right -- a visualization of electric field. light yellow color -- lower values, pale green color -- values at the edges of the boxes, blue color -- higher values))} 	
		\label{sch:rismviz}

	\end{figure}
	
\begin{figure}
		
		\includegraphics[width=5.0in]{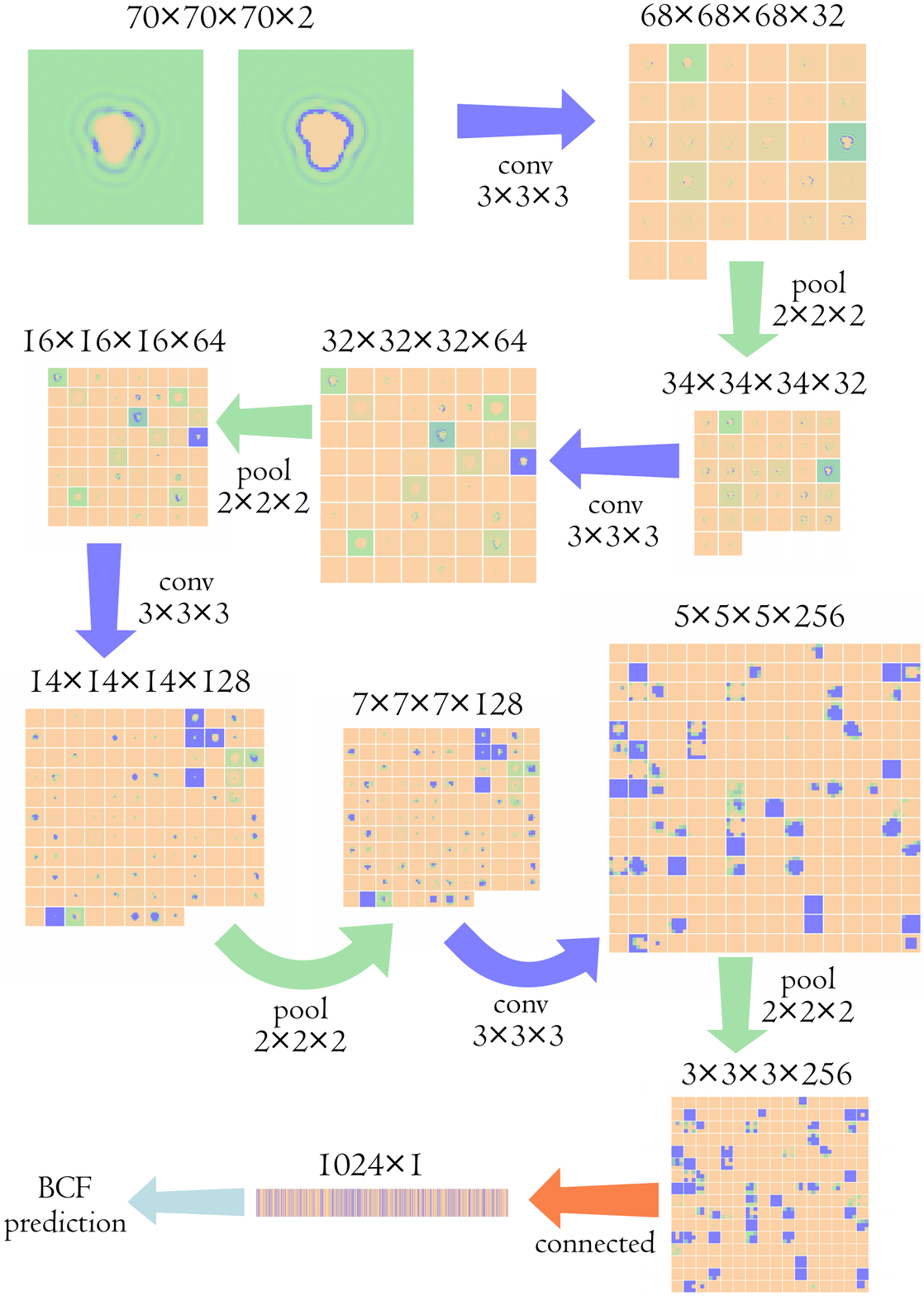}
		\caption{A schematic representation of \textit{ActivNet4} architecture with visualized 2D slices of feature maps on a trained network. Feature maps are colored using the same color scheme as in figure \ref{sch:rismviz}. Blue arrows labeled conv $\mathrm{N\times N \times N}$ denote a 3D convolution layer, green arrows labeled pool $\mathrm{N\times N \times N}$ denote 3D max-pulling layer, and red arrow labeled "connected" denotes a fully-connected layer. The figure is based on figure 4 from Ref. \citenum{Recent3Dconv2017}}
		\label{fig:net}
	\end{figure}

	\textit{3D Convolutional Neural Networks Modeling Procedure}
	We used framework chainer\cite{Chainer2015} to build networks for processing 3D data. The architecture of the network is schematically presented in figure \ref{fig:net} (a more standard representation is provided in table S1 in the Supporting Information). This architecture was optimal in terms of speed and the quality of the training models. This model has been called \textit{ActivNet4}, with four indicating the number of convolutional layers used.
	
	We trained this network on 3D fields using electric potentials and 3D-RISM results. Since we used water as a solvent in 3D-RISM calculations, we had two input channels (H and O atoms) in the 3D-RISM data, but only one in case of electric potentials.

	We used \textit{Parametric Rectified Linear Units}\cite{PreLu2015} as activation functions in the model, because they showed small improvement on the quality although it is possible to replace them with the common \textit{relu} activation function without noticeable lack of performance. To train \textit{ActivNet4}, we experimented with several optimizers: Stochastic gradient descent with momentum, \textit{Adam}\cite{Adam2014}, \textit{RMSprop}\cite{RMSprop2012}, and \textit{SMORMS3}\cite{SMORMS32015}. The best and stable convergence has been provided by \textit{SMORMS3} method. \textit{RMSprop} and \textit{Adam} have a good convergence ability, but the training process was less stable. Stochastic gradient descent has converged noticeably more slowly for the network. The parameters of the optimisers can be found in the Supporting Information. 
	
	The training and test processes were slightly different. At the training stage, each conformer of the molecule has been regarded independently from the other conformers. At the test stage, the prediction value for each conformer of the molecule has been calculated and the final result was the mean value for all conformers of the molecule. The performance of the model was estimated on the same test set that has been used in the original work to compare our model with the baseline. Further, we used 5-fold cross-validation (CV) technique to account the quality of the model in a more trusted way.
	The Neural networks have been trained using Nvidia K80 graphics cards and Nvidia GTX 1080 cards.
	
	\textit{Extreme Gradient Boosting modeling}
	To compare our 3D convolutional network with other machine learning approaches we built models using Extreme Gradient Boosting (XGBoost implementation\cite{XGBoost2016}) algorithm. This method has been proposed for use in cheminformatics\cite{XGBoostForQSAR2016} and can process very large datasets rapidly and efficiently. In this experiment, initially, we had to decrease the volume of each 3D cube from \textit{70x70x70} to \textit{17x17x17} by performing the average pooling operation with a kernel \textit{(4,4,4)}. Then, both oxygen and hydrogen channels have been flattened and stacked forming a vector of \textit{9826} values. These vectors served as the inputs for XGBoost algorithm. The application of the method to the test set has been performed in the same manner as in the neural network experiment.  We used the maximal number of trees = 100 and maximal depth of each tree = 6 to train the models, the other parameters have been set to default.
	
	\section{Results and Discussion} 
	
\begin{figure}
		\begin{centering}
		\includegraphics[width=3.25in]{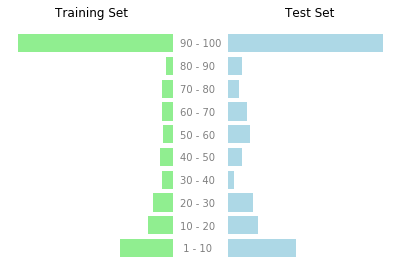}
		\end{centering}
		\caption{The distributions of the number of conformers for each molecule in the training and test sets}
		\label{fgr:confdistribution}
	\end{figure}

	Our main goal was to evaluate whether it is possible to predict biological property using a combination of solvation structure and machine learning. For this we took 670 molecules with known bioaccumulation constants and split them into a training (537 molecules) and test (133 molecules) sets. For each molecule we then generated a diverse set of conformers, using an earlier described procedure. The distribution of a number of conformers for both training and test set is shown in figure \ref{fgr:confdistribution}.
	
	Figure \ref{fgr:confdistribution} shows that about a quarter of the molecules in the training and test sets have less than 10 conformers (quite inflexible), while the rest consists of highly flexible molecules with 90-100 conformers. The distribution of the conformers is similar for the train and test sets.
	
	\begin{table}
		\caption{Accuracies of $\log_{10} \mathrm{BCF}$ predictions by different models. RMSE stands for root mean square error, MAE stands for mean absolute error and R denotes Pearson's correlation coefficient. }
		\label{table:activenet4results}
		%\rowcolors{2}{lightgray}{white}
		\begin{tabular}{l|c|c|c|c}
			\multicolumn{2}{c|}{\textbf{Model}}  & \textbf{RMSE} & \textbf{MAE} & \textbf{R\textsuperscript{2}}\\
			\hline 
			\multirow{2}{*}{US EPA (baseline)} & consensus model & 0.66  & 0.51 & 0.76 \\
			& single model & 0.68  & 0.64 & 0.74\\
			\hline
			\multirow{2}{*}{ActivNet4 (3D-RISM)} & training/test & 0.66  & \textbf{0.48} & \textbf{0.77} \\
			& 5-fold CV & 0.65 & 0.48 & 0.77 \\
			\hline
			\multirow{2}{*}{ActivNet4 (electric potential)} & training/test & 0.72  & 0.53 & 0.72 \\
			& 5-fold CV & 0.72 & 0.54 & 0.72 \\
			\hline
			\multirow{2}{*}{ActivNet4 (Lennard-Jones potential)} & training/test & 1.03 & 0.81 & 0.42 \\
			& 5-fold CV & 1.07 & 0.86 & 0.38 \\
			\hline
			\multirow{2}{*}{ActivNet4 (Lennard-Jones and electric
			potentials)} & training/test & 0.70 & 0.51 & 0.73 \\
			& 5-fold CV & 0.66 & 0.49 & 0.76 \\
			\hline
			\multirow{2}{*}{XGBoost (3D-RISM) } & training/test & 0.85  & 0.70 & 0.61 \\
			& 5-fold CV & 0.91 & 0.72 & 0.54 \\
		\end{tabular}
	\end{table}
	
	\begin{figure}
		\begin{centering}
		\includegraphics[width=3.25in]{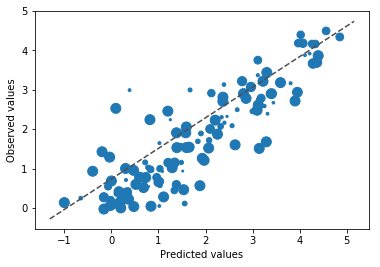}
		\end{centering}
		\caption{Correlation between observed and predicted values of $\log_{10} \mathrm{BCF}$. The size of the marker depends on the number of conformers of the molecule.}
		\label{fgr:obsvspred}
	\end{figure}
	
	The main result of the paper is presented in Table \ref{table:activenet4results}.
	As one can see, the ActivNet4 model has been capable of achieving a quality comparable to the ``consensus'' model provided by US EPA\cite{TESTEPA2016}. This result is noteworthy due to the fact that our model was based only on the 3D distribution of water molecules while the EPA's models used a large set of descriptors of varying nature. The comparison of the two models reveals that the analysis of solvent density distributions using 3D Convolutional neural networks may be useful for predicting biological properties. In addition, we demonstrate that the distribution can be correlated with the bioconcentration factor.

    We compared the results obtained by RISM, electric and Lennard-Jones potentials fileds. The quality of models on the base of RISM is slightly higher than ones built on electric field and notably higher than ones, based on Lennard-Jones potential. It can be interpreted, that taking into account solvation effects have an important role	in bioconcentration processes. The question of why the qualities of models on the base of Lennard-Jones potentials fields has less quality than the models based on electric potentials should be explained in the future researches.  

	One of the problems in our approach is related to the complexity of its set up. Indeed, both 3D-RISM calculations as well as 3D convolutional neural networks require some expertise to utilize. To check whether either of these procedures is necessary we also trained our network on electric potentials of molecules (thrid row in table \ref{table:activenet4results}) as well as used 3D-RISM obtained fields in combination with Extreme Gradient Boosting (XGBoost, fourth row of the table) algorithm. Both alternatives demonstrated worse results compared to the original, indicating that both 3D-RISM as well as CNN-s are necessary to achieve accurate results. Additionally, to address the difficulty of the codes we created a convenient script to simplify whole procedure, located on github \cite{Sosnin2017vga}.
	
	Another bottleneck of the proposed techniques is the size of the 3D fields. For a $70\times70\times70$  point 3D grid one has to spend a minimum of $ \SI{4}{\byte} \cdot70^3 = \SI{1372000}{\byte} $ (\SI{1.31}{\mega\byte}) to store it. 
	In the case of an n-site model of the solvent coupled with an m conformer representation of the solute we arrive to $ \SI{4}{\byte}\cdot 70^3 \cdot m \cdot n$ bytes necessary for each molecule. 

	To reduce the space requirements one can turn towards lossy compression methods for 3D fields. 3D autoencoders seem to be a good choice for this purpose. Another way is to develop of a method for generating 3D fields ``on-the-fly'' just before the training iteration. For instance, electric potentials used in this study were generated using quick GPU procedure that was essentially instantaneous (further details available in the supporting information).

\section{Conclusions} 
	The aim of the paper was to demonstrate that average solvent distribution in the neighbourhood of solutes can be used in combination with machine learning algorithms to predict properties that do not necessarily follow from the solvation structure alone. In order to achieve it we decided to focus on predicting the bioaccumulation factor using an approximate solvent density obtained using 3D-RISM method of integral equation theories. After training, the \textit{ActivNet4} (4-layer convolutional neural network) managed to predict $\log_{10}$BCF from water density distribution with RMSE=0.66. Although the model used relatively simple 3D descriptors, it managed to achieve prediction accuracies that are comparable to the state of the art models.
	
	Despite successful first results, the presented method requires further development. The first task that authors are working on right now is the application of the method to other molecular properties that are difficult to measure. Additionally, it is useful to explore possibilities of integrating solvation shell calculations and training steps to avoid storage limitations. Finally, given a clear physical meaning of the descriptors used in this study, it would be useful to explore precisely which molecular features significantly affect BCF. We hope to answer these and other questions in a follow-up article.
	
	The source code for the 3D fields generation is located  on Zenodo \doi{10.5281/zenodo.835526} and GitHub \href{https://github.com/sergsb/clever}{https://github.com/sergsb/clever}. It is distributed under Apache License 2.0.
	
	\section{Acknowledgements} 
	HPC calculations have been performed on Skoltech Pardus cluster. \textit{Instant JChem} was used for structure database management\cite{JChem2016}. Computations on GPU was carried out using computing resources of the federal collective usage center ``Complex for simulation and data processing for mega-science facilities'' at NRC ``Kurchatov Institute''.
	Authors acknowledgment Yermek Karpushev, Evgeny Burnaev, Alexey Zaytsev and David Palmer for fruitful discussions.

	\bibliographystyle{iopart-num}
	\bibliography{3dconv2017,zotero_export}
	
\end{document}

\section{Conformer generation}
	
	We used two sets: $C_{gen}$ and $C_{keep}$. $C_{gen}$ was a set of the generated conformers and $C_{keep}$ was a set with selected conformers. Initially, we calculated the energies of all conformers and select the lowest one $C_{low}$. Then we moved the $C_{low}$ conformer from $C_{gen}$ to $C_{keep}$. After that we estimated the \emph{RMSD} between each conformer in $C_{gen}$ and each one in $C_{keep}$ and moved from $C_{gen}$ to $C_{keep}$ only conformers with \emph{RMSD} more than predefined threshold (with threshold equal to \SI{0.3}{\angstrom}). Selected conformers have been saved as PDB files for future partial charges calculations and 3D RISM routines. Another selection approach, based on Taylor and Butina clustering algorithm\cite{ButinaCluster1999}, resulted in significantly worse results. Our experiments revealed that the standardization of the orientation of the conformers is a crucial step for future modeling. For standardizations, all conformers of each molecule have been orientated such the principal axes of the molecule have been aligned with the \emph{x,y,z} axes. 
	
\section{ActivNet4 architecture}

\begin{table}
	
	\rowcolors{2}{lightgray}{white}
	\begin{tabular}{l|c|c|c|c}
		\textbf{Layer} & \textbf{Kernel or pool size} & \textbf{Activation}\\
		\hline 
		conv1 & (3, 3, 3) & prelu \\
		pool1 & (2, 2, 2) & -- \\
		conv2 & (3, 3, 3) & prelu\\
		pool2 & (2, 2, 2) & -- \\
		conv3 & (3, 3, 3) & prelu\\
		pool3 & (2, 2, 2) & -- \\
		conv4 & (3, 3, 3) & prelu\\
		pool4 & (2, 2, 2) & -- \\
		\textbf{Dense layer} & \textbf{Neurons} & \textbf{Activation}\\
		\hline 
		fc1 & 1024 & prelu \\
		output & 1 & -- \\
	\end{tabular}
	\caption{The architecture of \textit{ActivNet4}. \emph{convX} denotes a 3D convolution layer, \emph{poolX} layer denotes a 3D max-pulling layer, and \emph{fcX} denotes a fully-connected layer, where \emph{X} is the sequential number of the layer.     
	}
	\label{table:activnet4structure}	
\end{table}

\section{Parameters of optimizers}
To train our networks we used the parameters listed below. For \emph{Adam} optimizer $ \alpha=0.001 $, $\beta_1=0.9 $, $\beta_2=0.999 $, $\epsilon=1e-08 $ parameters have been set. For \emph{RMSProp} we used \emph{learning rate} = 0.01, $ \alpha=0.001 $, $ \epsilon=1e-08 $ values. For Stochastic gradient descent we used \emph{learning rate} = 0.01, \emph{momentum} = 0.9. All other parameters were set to default.

\section{Comparison of calculation time on CPU and GPU}
	We compared the code for the electric potential calculations written for execution on CPU and GPU. We have performed CPU test on Intel(R) Core(TM) i7-6700 CPU with frequency up to 3.40GHz. We implemented the code in \emph{numpy} (CPU) and \emph{cupy} (a part of \emph{chainer} framework\cite{Chainer2015}) (GPU) versions. GPU tests have been performed on GeForce GTX 1080 card. We concluded that the GPU version is from 20 to 30 times faster than the CPU one, for the wide range of molecule sizes.
	
	\begin{figure}
		\includegraphics{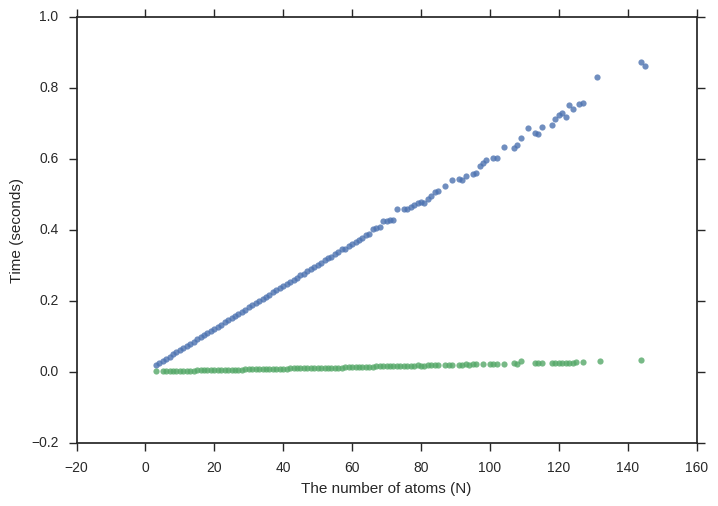}
		\caption{Comparison of calculation time on CPU (blue) and GPU (green) depending on the number of atoms in molecule (N). }
		\label{fgr:confvserr}
	\end{figure}

	\begin{figure}
		\includegraphics{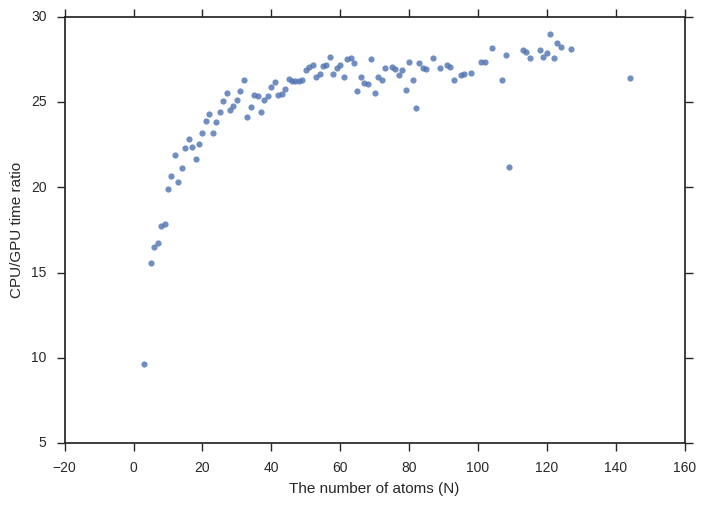}
		\caption{Ratio of calculation time on CPU to calculation time on GPU depending on the number of atoms in molecule (N).}
		\label{fgr:confvserr_gpu}
	\end{figure}

\bibliography{3dconv2017}